\documentstyle[12pt]{article}
\begin{document}
\large\rm
\renewcommand{\thesection}{\Roman{section}}
\renewcommand{\Re}{\mathop{\rm Re}\nolimits}
\renewcommand{\Im}{\mathop{\rm Im}\nolimits}
\renewcommand{\imath}{{i}}
\newcommand{\e}{\mathop{\it e}\nolimits}
\newcommand{\Sp}{\mathop{\rm Sp}\nolimits}
\newpage
\pagenumbering{arabic}
\thispagestyle{empty}
\begin{center}
M.V. Lomonosov Moscow State University\\
D.V. Skobeltsyn Institute of Nuclear Physics\\
\vspace{5cm}
{\LARGE\rm A FACTORIZATION OF THE\\\vspace{5mm} S--MATRIX
INTO JOST MATRICES}\\\vspace{1cm} {\large\rm A.F.~Krutov}\\{\large\it
Samara State University, 443011, Samara, Russia}\\\vspace{2mm}
{\large\rm D.I.~Muravyev, V.E.~Troitsky}\\
{\large\it Nuclear Physics Institute, Moscow State University, 119899,
Moscow, Russia}\\\vspace{1cm}
Preprint INP MSU-96-13/420\\\vspace{6cm}
MOSCOW 1996
\end{center}
\newpage\thispagestyle{empty}
\noindent
\begin{center}
{\bf abstract}\\
\end{center}

\noindent
An effective algebraic approach to $S$--matrix factorization
into Jost matrices
is developed in the
case of coupled channels.
The Jost matrix is given as a solution of boundary value
Riemann -- Hilbert problem.
A rational form is assumed for tangent of mixing angle, while
there is no limitations for approximation of phase
shifts.\\

\vspace{3cm}
\noindent
A.F. Krutov e-mail: krutov@ssu.samara.emnet.ru\\
D.I. Muravyev e-mail: muravyev@theory.npi.msu.su\\
V.E. Troitsky e-mail: troitsky@theory.npi.msu.su\\
\vspace{9.9cm}
\begin{center}
\copyright 1996, A.F. Krutov, D.I. Muravyev, V.E. Troitsky
\end{center}
\newpage
\section*{Introduction}
As it is well known, the Jost function plays an important role
in scattering theory \cite{New82}, and particularly in the
inverse scattering problem \cite{New89}--\cite{Tro94}.  One of
the methods of obtaining Jost function $F(k)$ is to solve
Riemann -- Hilbert problem for the half--plane.
In the frame of scattering theory the solution of this problem
is reduced to the \,$S$--matrix factorization in terms of Jost
matrices \cite{New82}.
The Jost matrix is also used to
obtain the solutions of nonlinear differential equations (see,
e.g. \cite{ZaM80}, \cite{DoE82}).

At present time a number of different
approaches to
the Jost matrix construction exists,
e.g. \cite{FuN56}--\cite{Coz93}.
However, while
the boundary
value problem solution
for one channel elastic scattering
can always be represented
explicitly in terms of
phase shift, this is not so for two--channel scattering.
In general case the Jost matrix can not be constructed
effectively.
It is known only that the matrix boundary value problem can be
reduced to a system of singular integral equations \cite{Vek67}.
While solving it
one needs to take into
account the following important fact.
Jost matrix arises at an intermediate stage in the investigation
of various problems, so its explicit form must be the
most adequate for analytical and numerical calculations.

In this connection,
one is interested to obtain the solution
avoiding
integral equations.

Let us note that the procedure of obtaining the Jost
matrix
can be effectively simplified if the \,$S$--matrix is factorized
previously ~\cite{Tro78}.
This enables one to write the result in a compact form, as can be
seen below from the concrete example.

In this paper we shall use the approximate form for mixing
parameter. Some of preliminary results were presented
in ~\cite{KiK93}.

This paper is organized as follows.
In Sec.~\ref{js2} the boundary value Riemann -- Hilbert problem
for \,$S$--matrix is formulated. The special features
of the problem in matrix case are discussed in
Sec.~\ref{js3}. Our algebraic approach consists of three steps.
In Sec.~\ref{js4} the first step \,$S$--matrix
factorization --- prefactorization is realized.
The "correcting matrix" concept
needed
to solve the boundary
value problem is discussed
in Sec.~\ref{js5}. In Sec.~\ref{js6}
the final result for Jost matrix is obtained.
The effectiveness of our approach is demonstrated
in Sec.~\ref{js7} for
the well--known effective radius approximation.

\section{The formulation of the problem}\label{js2}

The general mathematical formulation of the problem is
the following. One needs to find the piecewise holomorphic matrix
$F(k)$ (i.e.  its matrix elements are
piecewise holomorphic functions) in the upper and lower
half--planes. Boundary values of $F(k)$ satisfy the
following condition on the real axis:
\begin{equation}
\label{j21} F_+(k) = S^{-1}(k)\,F_-(k)\>,\quad
\hbox{Im}\,k = 0\>,\quad -\infty\><\,k\,<\>\infty
\end{equation}
Here $F_+(k)$ is a matrix holomorphic in the upper half--plane,
$F_-(k)$ --- in
the lower one, $k$ is momentum in the center--of--mass system,
$S(k)$ is the
scattering matrix which is non--singular:
\begin{equation}\label{j22}
   \det S(k)\ne 0\,,\quad \Im k=0
\end{equation}
and H\"older condition is valid for its matrix elements
$s_{ij}(k)\,,\,i,j\!=\!1,2$ on the real axis
\begin{equation}\label{j23}
   |s_{ij}(k_1)-s_{ij}(k_2)|\le A|k_1-k_2|^\mu,
   A>0\,,0<\mu\le1\;, i,j=1,2\,.
\end{equation}
Standard conditions are imposed on the scattering matrix
$S(k)$ and on the $F(k)$ matrix
(the boundary values of the matrix \, $F(k)$\,
 are the Jost matrices)~\cite{New82}:
\begin{equation}\label{j24}
   S^\dagger (k)=S^{-1}(k)=S^*(k)=S(-k)\,,\quad \Im k=0\,,
\end{equation}
\begin{equation}\label{j25}
   \lim_{k\to \pm \infty}S(k)=I\,,
\end{equation}
\begin{equation}
F_+^*(k) = F_+(-k) = F_-(k)\>,\quad
\hbox{Im}\,k = 0\>,
\label{jost}
\end{equation}
$I$ is the unity matrix.
As usual the physical
formulation of the problem requires, in addition,
the asymptotic condition
\begin{equation}\label{j26}
   \lim_{k\to\pm\infty}F_+(k)=I
\end{equation}
and the condition
\begin{equation}\label{j27}
   \det F_+(\imath\kappa_j)=0\,,\quad \kappa_j>0\,,j\!=\!1,2,\dots ,m\,,\quad
   m<+\infty\,,
\end{equation}
to be satisfied
($m$ is the number of bound states).

\section{Boundary value problem in the\protect\\ matrix case }\label{js3}

In
the scalar (non--matrix) case of uncoupled partial waves the
solution of the problem formulated above
can be given
explicitly \cite{Gak66}:  \begin{equation}\label{j31} F_\pm
   (k)=\Pi_\pm (k)\exp\left(\frac{1}{2\pi\imath}
				    \int\limits^{+\infty}_{-\infty}
                                    \frac{\ln \left( S^{-1}(k)
                                                      \Pi^2_-(k)
                                               \right)}{k'-k\mp\imath 0}
                                    \,dk'
                             \right)\,,
\end{equation}
\begin{equation}\label{j32}
   \Pi_\pm (k)=\prod^m_{j=1}\frac{k\mp\imath\kappa_j}{k\pm
                                             \imath\kappa_j}\,,
\end{equation}
\begin{equation}\label{j33}
   \Pi_\pm (k)\equiv1\,,\quad m=0\,.
\end{equation}
However, in general case the matrices given by the
Eqs.(\ref{j31})--(\ref{j33})
do not satisfy the boundary value condition (\ref{j21})\,.
In fact, using the well--known equation
$$
   \frac{1}{k'-k\mp\imath0}=\frac{1}{k'-k}\pm\imath\pi\delta (k'-k)
$$
and substituting (\ref{j31}) in (\ref{j21}),
one obtains
$$
   \e^{\frac{1}{2}g(k)+h(k)}=\e^{g(k)}e^{-\frac{1}{2}g(k)+h(k)}\,,
$$
with
$$
   g(k)=\ln(S^{-1}(k)\Pi^2_-(k))\,,\quad
h(k)=P.V.\int\limits^{+\infty}_{-\infty}\frac{g(k')}{k'-k}\,dk',
$$
P.V. stays for the
principal value of the integral.
So if we define the matrices $A$ and $B$ by
$$
   A=g(k)\,,\quad B=-\frac{1}{2}\,g(k)+h(k)
$$
then the rule
\begin{equation}\label{j34}
   \e^{A+B}=\e^A\e^B
\end{equation}
is to be fulfilled.
The Eq. (\ref{j34}) is valid only if
\begin{equation}\label{j35}
   S(k_1)\,S(k_2)=S(k_2)\,S(k_1)\,,\quad k_1\neq k_2\,.
\end{equation}
The condition (\ref{j35}) means, from the physical
point of view, that the potential in Schr\"oedinger equation is
central. If the interaction is of tensor character, then
the condition (\ref{j35}) and, consequently,
the Eq. (\ref{j34}), are not fulfilled. This means that
\,(\ref{j31})\, can not be used for solving the
boundary value problem of Sec.~\ref{js2}.

To construct the Jost matrix in this case let us begin with
the first stage ---
preliminary factorization of the \,$S$--matrix
--- "prefactorization".

\section{The \,$\bf S$-matrix prefactorization}\label{js4}

Now the problem of finding of Jost matrix in the coupled channels
case is reduced to
the boundary value problem (\ref{j21}) with conditions
(\ref{j22})--(\ref{j27}).
$S$--matrix can be diagonalized by the use of an orthogonal
transformation \begin{equation}\label{j41}
   S(k)=U(k^2)\left(\begin{array}{cc}
                       \e^{\imath 2\delta_1(k)}&0\\
                       0&e^{\imath 2\delta_2(k)}
                    \end{array}
              \right)
        U^{-1}(k^2)\,,\quad\Im k=0\,,
\end{equation}
where \, $\delta_j(k)\,, j\!=\!1,2$\,
are the real--valued phase shifts,
$$
\delta_j(-k)=\,-\,\delta_j(k)\,,\quad \delta
(\pm\infty)=0\,,\quad j\!=\!1,2\,.
$$
The matrix \, $U(k^2)$\,
as usually can be written in the form:
$$
U(k^2)=\left( \begin{array}{rc} \cos\varepsilon
(k^2)&\sin\varepsilon (k^2)\\
-\sin\varepsilon (k^2)&\cos\varepsilon (k^2)
\end{array}
\right)
$$
where $\varepsilon (k^2)$ is the real--valued
mixing angle.
We shall approximate the tangent of mixing angle
$\varepsilon(k^2)$  by a rational function:

\begin{equation}\label{j42}
   \tan\varepsilon (k^2)=\frac{P_M(k^2)}{Q_N(k^2)}\,.
\end{equation}
Here $P_M(k^2)$ and $Q_N(k^2)$ are
polynomials in $k$ of the degree $2M$ and $2N$, respectively.
Without loss of generality one can suppose that
the polynomials
have no zeros simultaneously.
The diagonal part of the $S$--matrix
(\ref{j41}) can be factorized:
   \begin{equation}\label{j43}
   \left(\begin{array}{cc}
            \e^{2\imath\delta_1(k)}&0\\
            0&e^{2\imath\delta_2(k)}
         \end{array}
   \right)=
   \left(\begin{array}{cc}
            f_{1-}(k)&0\\
            0&f_{2-}(k)
         \end{array}
   \right)
   \left(\begin{array}{cc}
            f^{-1}_{1+}(k)&0\\
            0&f^{-1}_{2+}(k)
         \end{array}
   \right)\,,
\end{equation}
where $f_{j\pm}(k), j\!=\!1,2$\,
are the solutions 
$$
   f_{j+}(k)=\e^{-2\imath\delta_j(k)}\,f_{j-}(k)\,,\quad
j\!=\!1,2, $$
of scalar boundary  value problems
given by the Eqs.(\ref{j31})--(\ref{j33}).
Let us note, that if the number of bound states in the first
channel (for $j\!=\!1$) is equal to \,$m_1$\, then one must
change \,$m$\, for \,$m_1$\, in (\ref{j31})--(\ref{j33}).
Similarly one has to change \,$m$\, for \,$m_2$\, for
\,$j\!=\!2$\,; $m_1\!+m_2=m$\,.
Taking into
account \,(\ref{j42})\,,\,(\ref{j43})\, the problem \,
(\ref{j21})\, can be written in the form:
\begin{equation}\label{j44}
   F^{(0)}_+(k)=S^{-1}(k)\, F^{(0)}_-(k)\,
\end{equation}
with
\begin{equation}\label{j45}
   F^{(0)}_\pm (k)=\left(
                         \begin{array}{rc}
                            Q_N(k^2)&P_M(k^2)\\
                           -P_M(k^2)&Q_N(k^2)
                         \end{array}\,
                   \right)
                   \left(\begin{array}{cc}
                          f_{1 \pm}(k)&0\\
                          0&f_{2 \pm}(k)\,
                         \end{array}
                   \right).
\end{equation}
So the preliminary factorization of \,$S$--matrix is realized.

\section{The analytical properties of\protect\\
the matrix
\,$F^{(0)}_+(k)$}\label{js5}

The matrix
\, $F^{(0)}_+(k)$\, has a number of properties which are
characteristic for the Jost matrix. As it follows from
Eq. (\ref{j45})\  $F^{(0)}_+(k)$\, is analytical in
the upper
half--plane and the following equations take place:
\begin{equation}\label{j51}
   F^{(0)*}_+(k)=F^{(0)}_+(-k)=F^{(0)}_-(k)\,,\quad \Im k=0\,.
\end{equation}
Nevertheless, $F^{(0)}_+(k)$\,
is not a solution to the boundary value problem of Sec.~\ref{js2}
because: 1) the condition \,(\ref{j26})\, is not satisfied;
2) $\det F^{(0)}_+(k)$ has
nonphysical extra zeros
originated by the roots of the equations:
\begin{equation}\label{j52}
   Q_N(k^2)+\imath P_M(k^2)=0\,,
\end{equation}
\begin{equation}\label{j53}
   Q_N(k^2)-\imath P_M(k^2)=0\,.
\end{equation}
Using (\ref{j45}) one can write
$$
\det F^{(0)}_+(k)=\left(
                        Q^2_N(k^2)+P^2_M(k^2)
                  \right) f_{1+}(k)\,f_{2+}(k)\,
$$
so that the condition (\ref{j27})\, is
not satisfied.

Let us emphasize one particularly important
feature of the roots of
the Eqs.\,(\ref{j52})--(\ref{j53})\,. The polynomials are
even functions of $k$ and are not  equal to zeros simultaneously
(see Sec.~\ref{js4})\,, so that the Eqs. \,(\ref{j52})\,,\,(\ref{j53})\, 
have no roots if
\,$\Im k\!=\!0$\,, or if $\Re k\!=\!0$.
Let us also notice
that for real \,$k$\, the Eqs.(\ref{j52})\,,\,(\ref{j53})\,
are complex conjugate. Now we can denote
the roots of the Eqs.\,(\ref{j52})\,
and \,(\ref{j53})\,
by
\,$\pm\imath\lambda_j,
\pm\imath\lambda^*_j,$ $j\!=\!1,2,\dots ,L,\,
L=\max\{M\,,N\}$\,.
For definiteness let us
suppose
\,$\Im (\imath\lambda_j)>0\,, j\!=\!1,2,\dots ,L$\,.
It is easy to see
that the positions of all roots are symmetrical
relatively to
real and imaginary
axes.
We shall refer to these roots as to nonphysical zeros of the
\,$F^{(0)}_+(k)$\,--matrix determinant.
The symmetry described in this Sec. is of great importance for
the solution of the boundary value problem.

Let us consider first the case of simple zeros.
In order to obtain the Jost matrix, one needs only to
remove
the nonphysical zeros from the \, $\det F^{(0)}_+(k)$\,,
but in such a way that
the analytical properties of the matrix
\,$F^{(0)}_+(k)$\, remain unchanged
and the condition (\ref{j26}) is fulfilled.
Let us multiply the boundary value condition
\, (\ref{j44})\,
from the right by a rational real matrix \, $W(k^2)$
$$
   F^{(0)}_+(k)\,W(k^2)=S^{-1}(k)\,F^{(0)}_-(k)\,W(k^2)\,.
$$
If the matrix \, $W(k^2)$\, is such, that conditions
on \, $F^{(0)}_+(k)$\, mentioned above
are fulfilled, then the Jost matrix  $F_+(k)$
is of the form
\begin{equation}\label{j54}
   F_+(k)=F^{(0)}_+(k)\,W(k^2)\,.
\end{equation}
We will refer to
the matrix \, $W(k^2)$\, as the correcting matrix.
The method of obtaining the explicit form of \, $W(k^2)$\,
is based on the ideas of \cite{FuN56}, \cite{Tro78},
\cite{Che56}.

\section{The construction of the
correcting matrix \,$W(k^2)$}\label{js6}

Let us make use of the symmetry property of the nonphysical
zeros (Sec.~\ref{js5}\,) and let us fix four of them:
\,$\pm\imath\lambda_1,\pm\imath\lambda^*_1$\,.
We shall now remove
these nonphysical zeros from \, $\det F^{(0)}_+(k)$\,.
To do this let us consider first the matrix
\begin{equation}\label{j61}
   W_1(k^2)=P_1\frac{1}{(k-\imath\lambda_1)(k+\imath\lambda_1)}+
(I-P_1)\frac{1}{(k-\imath\lambda^*_1)(k+\imath\lambda^*_1)},
\end{equation}
where\, $P_1$\, is a
projection matrix
\begin{equation}\label{j62}
   P^{ 2}_1=P_1.
\end{equation}
Using the fact that (\ref{j62}) implies
$$
   \det P_1=0\,,\quad \Sp P_1=1\,,
$$
it is easy to check that
\begin{equation}\label{j63}
   \det W_1(k^2)=\frac{1}{(k+\lambda^2_1)(k+\lambda^{2*}_1)}\,.
\end{equation}
Now we can obtain
a new matrix:
$$
   F^{(1)}_+(k)=F^{(0)}_+(k)\,W_1(k^2)\,.
$$
As can be seen from Eq.(\ref{j63}) the
$\det F^{(1)}_+(k)$ has no zeros at the points
$k=\pm\imath\lambda_1$,\, $\pm\imath\lambda^*_1$.

Now we need to verify
the analyticity of \,$F^{(1)}_+(k)$\, in the upper half--plane
and the properties analogous
to (\ref{j51}).
This second condition will be satisfied if
\begin{equation}\label{j64}
   P^*_1=I-P_1\,.
\end{equation}
The explicit form of\, $P_1$\, must be chosen in such
a way
that the matrix\, $F^{(1)}_+(k)$\, will have no
poles
at the points \, $k=\imath\lambda_1, \,\imath\lambda^*_1$\,.
Taking into account the Eqs.
\,(\ref{j62})\,,\,(\ref{j64})\,,
we can define
\begin{equation}\label{j65}
   P_1=\frac{1}{\Sp Y_1}\,Y_1\,,
\end{equation}
with
\begin{equation}\label{j66}
   Y_1=\left(
            \begin{array}{cc}
               f_{2+}(\imath\lambda_1)&0\\
               0&f_{1+}(\imath\lambda_1)
            \end{array}
       \right)
       \left(
            \begin{array}{cr}
               1&-\imath\\
               \imath&1
            \end{array}
       \right)
       \left(
            \begin{array}{cc}
               f^*_{1+}(\imath\lambda_1)&0\\
               0&f^*_{2+}(\imath\lambda_1)
            \end{array}
       \right).
\end{equation}
Now $F^{(1)}_+(k)$ has no zeros in the upper half--plane.
To check
this fact it is sufficient to show that the following equations
are satisfied:
\begin{equation}\label{j67}
   F^{(1)}_+(\imath\lambda_1)\,P_1=0\,,
\end{equation}
\begin{equation}\label{j68}
   F^{(1)}_+(\imath\lambda^*_1)\,(I-P_1)=0\,.
\end{equation}
This can be done using the explicit form of
$F^{(1)}_+(k)$ and $ P_1$\,
( (\ref{j45}) and (\ref{j65})\,,\,(\ref{j66}),
respectively) and the fact that
$k=\imath\lambda_1,\, \imath\lambda^*_1$
are the roots of the Eqs.
(\ref{j52})\,,\,(\ref{j53}).

After removing
of $4(n-1), n \le L$ nonphysical zeros we obtain for
$F^{(n)}_+(k)$:
\begin{equation}\label{j69}
   F^{(n)}_+(k)=F^{(n-1)}_+(k)\,W_n(k^2)\,,
\end{equation}
\begin{equation}\label{j610}
   W_n(k^2)=P_n\frac{1}{(k-\imath\lambda_n)(k+\imath\lambda_n)}+
(I-P_n)\frac{1}{(k-\imath\lambda^*_n)(k+\imath\lambda^*_n)},
\end{equation}
where \,$P_n$\, has properties analogous
to the Eqs. (\ref{j62}), (\ref{j64})
and is of the form
\begin{equation}\label{j611}
   P_n=\frac{1}{\Sp Y_n}\,Y_n\,,
\end{equation}
\begin{eqnarray}
      Y_n=
          W^{-1}_{n-1}(-\lambda^2_n)\times\cdots\times
          W^{-1}_{1}(-\lambda^2_n)\phantom{-----------}\label{j612}\\
          \phantom{x}\nonumber\\
          \times\left(
                      \begin{array}{cc}
                         f_{2+}(\imath\lambda_1)&0\\
                         0&f_{1+}(\imath\lambda_1)
                      \end{array}
                \right)
                \left(
                      \begin{array}{cr}
                         1&-\imath\\
                         \imath&1
                      \end{array}
                \right)
                \left(
                      \begin{array}{cc}
                         f^*_{1+}(\imath\lambda_1)&0\\
                         0&f^*_{2+}(\imath\lambda_1)
                      \end{array}
                \right)\nonumber\\
                \phantom{x}\nonumber\\
                \phantom{------------}
                \times [W_{1}(-\lambda^2_n)\times\cdots\times
                W_{n-1}(-\lambda^2_n)]^*\,.\nonumber
\end{eqnarray}
It is easy to show that
$\det F^{(n)}_+(k)$ has no
nonphysical zeros
\,$k\!=\!\imath\lambda_j$,\,$\imath\lambda^*_j$,\,
$j\!=\!1,2,\dots,n$,\,
and the matrix $F^{(n)}_+(k)$ is
analytical in the upper half--plane.

When all nonphysical zeros are removed and
$L\, (L=\max\{M,N\})$\,
projection matrices are
obtained,
then taking into account the Eq. (\ref{j26})
one has for the correcting matrix:
\begin{equation}\label{j613}
   W(k^2)=W_1(k^2)\times\cdots\times W_L(k^2)\,C\,,
\end{equation}
\begin{equation}\label{j614}
   C=\cases{
            \phantom{---}
            \frac{1}{q_N}\,I\,,&$N>M$\,,\cr
            \frac{1}{p_M}\left(
                               \begin{array}{cr}
                                  0&-1\\
                                  1&0
                               \end{array}
                         \right),&
            $N<M$\,,\cr
            \phantom{1}
            \left(
                  \begin{array}{cc}
                     q_N&p_M\\
                    -p_M&q_N
                  \end{array}
            \right)^{-1},&$N=M$\,,\cr
           }
\end{equation}
where\, $p_M,q_N$\, are the highest degree coefficients
in the polynomials \, $P_M(k^2)$, \, $Q_N(k^2)$,
respectively.  The
Eqs.\,(\ref{j54})\,,\,(\ref{j61})--(\ref{j614})\, give the
solution of the boundary value problem of Sec.~\ref{js2} in the
case of the approximation \,(\ref{j42})\, for the mixing
angle.

The boundary value problem of Sec.~\ref{js2}\,
can be solved
in the case of
multiple
nonphysical zeros, too.
However, in this case, one cannot write simple explicit
equations for the projection matrix, although the idea of the
construction remains to be the same.
Let us notice, however, that, as was mentioned above, Jost
matrices are used usually as an intermediate step during the
solution of physical problems, and it is very useful to have
definite simple form of them. For this purpose it is more
convenient to have simple zeros. This is always possible to be
done remaining within the limits of the experimental error, 
keeping in mind that the polynomials $P_M(k^2)$, $Q_N(k^2)$ give 
an approximation for $\tan\varepsilon (k^2)$.

\section{The example}\label{js7}

As an example let us consider the elastic two--channel
neutron--proton scattering with the phase shifts
\, $^3S_1\,,\,^3D_1$\, in the effective radius
approximation. In this case the scattering phase shifts and
mixing angle are of the form:
$$
   \e^{\imath 2\delta_S(k)}=\frac{(k+\imath\kappa)(k+\imath\varphi)}
                                {(k-\imath\kappa)(k-\imath\varphi)}\,,\quad
   \e^{\imath 2\delta_D(k)}=1\,,\quad
   \tan\varepsilon (k^2)=\frac{k^2}{2\chi^2}\,,
$$
where \, $k\!=\!\imath\kappa$\, is the bound state point,
so that at \,$m\!=\!1$\, the condition\, (\ref{j27})\,
is to be taken into account.

The $S$--matrix is a rational function and
has the form:
$$
   S(k)=\frac{1}{k^4+4\chi^4}
        \left(
              \begin{array}{cc}
                 2\chi^2&k^2\\
                 -k^2&2\chi^2
              \end{array}
        \right)
        \left(
              \begin{array}{cc}
                 \frac{(k+\imath\kappa)(k+\imath\varphi)}
                      {(k-\imath\kappa)(k-\imath\varphi)}&0\\
                 0&1
              \end{array}
        \right)
        \left(
              \begin{array}{cc}
                 2\chi^2&-k^2\\
                 k^2&2\chi^2
              \end{array}
        \right)\,.
$$
Let us perform
the $S$--matrix prefactorization in terms of matrices
$F^{(0)}_\pm (k)$ (see
(\ref{j44}), (\ref{j45}))
$$
F^{(0)}_\pm (k)=\left(
 \begin{array}{cc}
                            2\chi^2&k^2\\
                            -k^2&2\chi^2
                         \end{array}
                   \right)
                   \left(
                         \begin{array}{cc}
                            \frac{k\mp\imath\kappa}
                                 {k\pm\imath\varphi}&0\\
                            0&1
                         \end{array}
                   \right)\,.
$$
In our case the polynomials
$P_M(k^2), Q_N(k^2)$ (see (\ref{j42}))
are of the form:
$$
   P_M(k^2)=k^2\,,\quad Q_N(k^2)=2\chi^2\,,
$$
with\, $M\!=1,N\!=0$\,. The number of projection operators
$P_j\,,\,j\!=1,2,\dots $
is $L\,,\,L=\max\{M,N\}$\, (see Sec.~\ref{js6})\,.
In our case
$L\!=1$,
so to obtain the Jost matrix we need to construct
only one projection matrix $P_1$.
The nonphysical zeros of\, $\det F^{(0)}_+(k)$\,
(the roots of Eqs. (\ref{j52})\,,\,(\ref{j53}))
are \, $k\!=\!\pm\lambda_1,\pm\lambda^*_1$\,,
with \, $\lambda\!=\!\chi (1-\imath )$\,.

Following (\ref{j65}), (\ref{j66}) it is easy to
obtain the projection matrix
$P_1$ in the form:
$$
P_1=\frac{1}{2\eta} \left(
\begin{array}{cc}
               \eta+\imath\chi (\varphi +\kappa)&
               -\imath(\chi^2+(\chi+\varphi )^2)\\
               \imath(\chi^2+(\chi-\kappa )^2)&
               \eta-\imath\chi (\varphi +\kappa)
            \end{array}
       \right)\,,
$$
$$
   \eta =\chi^2+(\chi +\varphi)(\chi -\kappa)\,.
$$
Using the condition (\ref{j26}) one has for the
Jost matrix:
$$
\begin{array}{l}
F_+(k)=\left( \begin{array}{cc} 2\chi^2&k^2\\
   -k^2&2\chi^2
                \end{array}
          \right)
          \left(
                \begin{array}{cc}
                   \frac{k-\imath\kappa}
                        {k+\imath\varphi}&0\\
                   0&1
                \end{array}
          \right)\phantom{-------------}\\
          \phantom{------------} 
          \times
          \left[
               P_1\frac{1}{k^2+\lambda^2_1}+
              (I-P_1)\frac{1}{k^2+\lambda^{2*}_1}
          \right]
          \left(
               \begin{array}{cr}
                  0&-1\\
                  1&0
               \end{array}
          \right)\,.
\end{array}
$$
Instead of the $P_1$ matrix  we can use the $P$ matrix 
defined by
$$
   P_1=\left(
             \begin{array}{cc}
                0&1\\
                -1&0
             \end{array}
       \right)
       P
       \left(
             \begin{array}{cr}
                0&-1\\
                1&0
             \end{array}
       \right).
$$
Now
\begin{equation}\label{j71}
   F_+(k)=\left(
                \begin{array}{cr}
                   k^2&-2\chi^2\frac{k-\imath\kappa}
                                    {k+\imath\varphi}\\
                   2\chi^2&k^2\frac{k-\imath\kappa}
                                   {k+\imath\varphi}
                \end{array}
          \right)
          \left[
               P\frac{1}{k^2+\lambda^2_1}+
              (I-P)\frac{1}{k^2+\lambda^{2*}_1}
          \right]\,.
\end{equation}
This simple case considered as an example was investigated in
\cite{NeF57} and one can show that\, (\ref{j71})\,
is nothing else that the well known result of \cite{NeF57}
for the Jost matrix, but in more compact form. Our projection
matrix $P$ is just $P_3$ --- one of the projection matrices of
\cite{NeF57}. However, the $S$--matrix prefactorization
and the use of the symmetry property for nonphysical zeros make
our algebraic method more effective and allows one to apply it
in more complicated cases.

\section{Conclusion}\label{js8}

We have developed an effective algebraic method of the explicit
factorization of the $S$--matrix in terms of the Jost matrices in
the case of tensor potential. The only approximation used is the
rational form for the mixing angle tangent. One of the main
advantages of the method is the fact that the number of
projection matrices involved does not depend on the phase shifts
approximations.  The obtained Jost matrices have simple compact
explicit form and can be used in a number of physical situations
(see e.g. \cite{GeK94}).

Our result, as a mathematical result, is of
independent interest by its own
rights, too. We have enlarged the class of unitary matrices for
which the solution of the Riemann -- Hilbert boundary value
problem can be obtained in a way avoiding systems of singular
integral equations.

\end{document}